\documentclass{article}
\usepackage{spconfa4,amsmath,graphicx}

\renewcommand{\k}{{[k]}}

\newcommand{\y}{{\bf y}}
\newcommand{\z}{{\bf z}}
\newcommand{\x}{{\bf x}}
\newcommand{\s}{{\bf s}}
\newcommand{\g}{{\bf g}}

\renewcommand{\a}{{\bf a}}
\newcommand{\w}{{\bf w}}

\title{Informed FastICA: Semi-Blind Minimum Variance Distortionless Beamformer}
\name{Zbyn\v{e}k Koldovsk\'{y}$^1$, Ji\v{r}\'i M\'alek$^1$, Jaroslav \v{C}mejla$^1$, and Stephen O'Regan$^2$\thanks{This work was supported by the Department of the Navy, Office of Naval Research Global, through Project No.~N62909-23-1-2084.}}
\address{$^1$Acoustic Signal Analysis and Processing Group, 
Technical University of Liberec, Czech Republic\\
$^2$Naval Surface Warfare Center Carderock Division, West Bethesda, Maryland, USA}
\begin{document}
\ninept
\maketitle
\begin{abstract}
Non-Gaussianity-based Independent Vector Extraction leads to the famous one-unit FastICA/FastIVA algorithm when the likelihood function is optimized using an approximate Newton-Raphson algorithm under the orthogonality constraint. In this paper, we replace the constraint with the analytic form of the minimum variance distortionless beamformer (MVDR), by which a semi-blind variant of FastICA/FastIVA is obtained. The side information here is provided by a weighted covariance matrix replacing the noise covariance matrix, the estimation of which is a frequent goal of neural beamformers. The algorithm thus provides an intuitive connection between model-based blind extraction and learning-based extraction. The algorithm is tested in simulations and speaker ID-guided speaker extraction, showing fast convergence and promising performance.
\end{abstract}

\section{Introduction}
\label{sec:intro}

We address the problem of extracting the source of interest (SOI) from a mixture of signals observed by multiple sensors. We particularly focus on speaker extraction from noisy multi-microphone recordings. Solutions to this problem can be achieved by beamforming or blind methods. On the one hand, beamforming offers optimal filters \cite{vantrees2002}. However, these can only be used if we know the key SOI or noise statistics, the precise estimation of which is as difficult a task as the problem itself. On the other hand, blind methods perform extraction based only on general assumptions such as the independence of SOI and other signals, as in Independent Vector Extraction (IVE) \cite{ASSSEbook2018,koldovsky2019TSP,scheibler2019overiva}. However, blind methods have limited accuracy and are burdened by uncertainties.

Both methodologies have, therefore, been adapted in various ways. Most recently, beamforming has been combined with deep neural networks that are trained to estimate the necessary SOI/noise covariance matrices \cite{heymann2016,wang2018}. Blind methods are modified so that they can use side information to improve their performance, which is referred to as semi-blind, guided or informed methods. For example, in speaker extraction, the knowledge of the speaker's location is often used for constraining algorithms to enforce the extraction of the desired speaker \cite{parra2002,khan2015,brendel2020}. There is also a significant class of methods that exploit information about the SOI activity, such as the variance profile of the speech or speaker's ID indicator \cite{onotakuma2012,loc:bse:defl}. Although semi-blind methods form a diverse class of approaches, we observe in the literature that in many cases, a weighted covariance matrix of the input data plays a role, where the weights are computed using side information \cite{cho2019,nakatani2019,hiroe2022,jansky2022}. Typically, the intention is that the weighted covariance matrix replaces the unknown noise/SOI covariance matrix: Semi-blind and beamforming methods meet here \cite{yamaoka2021,nakatani2022}.

In this paper, we propose an intuitive modification of the famous FastICA/FastIVE algorithm \cite{hyvarinen1999,lee2007fast} for semi-blind IVE. The original algorithm is derived using a fast second-order optimization of the likelihood function under the orthogonal constraint (OC). The OC enforces zero sample covariance between the estimated SOI and the other signals, which comes from the assumption of their independence. Analytically, the OC is equivalent to the minimum power distortionless (MPDR) beamformer, where the input-data covariance matrix and the mixing vector representing the relative acoustic transfer function (RTF) between the microphones and the SOI, play a role. The main idea here is to replace the covariance matrix in MPDR with its weighted counterpart, by which MPDR is replaced by an approximate MVDR (minimum variance distortionless beamformer), provided that the weighted covariance matrix approximates the noise covariance matrix. The weights are assumed to be provided through side information, which makes the algorithm semi-blind. Since MVDR is known to be less sensitive to the error in the mixing vector than MPDR, better global convergence to SOI can be expected. In simulations, we confirm this property, including improved accuracy, depending on how accurate the side information is. We also validate the algorithm on a speaker extraction task guided by deep training-based speaker identification.

The paper is organized as follows: The following section formulates the problem and recalls the MPDR and MVDR beamformers, as well as the basis for solving the problem using IVE. Section III describes the proposed algorithm and reveals its relations to the original FastICA. Section IV shows the results of the simulations and the speaker extraction experiment. Section V concludes the article.

\section{Problem Formulation}
\label{sec:problem}

\subsection{Observation and extraction model}
We consider $K$ mixtures of signals each observed through $d$ sensors (microphones). The $n$th sample of the $k$th observed mixture is represented by a $d\times 1$ vector $\x^\k(n)$ which contains the contribution of the source of interest (SOI) according to 
\begin{equation}\label{BSEmodel}
    \x^\k(n)=\a^\k s^\k(n) + \y^\k(n),
\end{equation}
where $n=1,\dots,N$, $k=1,\dots,K$, $\a^\k$ is the mixing vector of the SOI containing weights with which the SOI contributes to the sensors, and 
$\y^\k(n)$ contains the other signals in the mixture. The extraction model assumes a separating vector $\w^\k$ such that 
\begin{equation}\label{extraction}
    s^\k(n)\approx(\w^\k)^H\x^\k(n)
\end{equation}
in an exact or as close as possible sense. 
In the frequency-domain speaker extraction problem, $k$ and $n$ play the role of the frequency and frame index of the short-term Fourier transform, respectively. For simplicity, we omit writing the argument $n$ further in the paper.

\subsection{Optimum Beamformers}
Array processing theory introduces optimum beamformers that seek $\w^\k$ as a solution to an optimization problem \cite{vantrees2002}. In the context of this paper, we will consider two well-known beamformers. Minimum variance distortionless beamformer (MVDR) is the solution of 
\begin{equation}
    \w_{\rm MVDR}^\k=\arg\min_{\w} \w^H{\bf C}_{\bf y}^\k\w\quad\text{s.t.}\quad \w^H\a^\k=1,
\end{equation}
and minimum power distortionless beamformer (MPDR) is given by
\begin{equation}
    \w_{\rm MPDR}^\k=\arg\min_{\w} \w^H{\bf C}_{\bf x}^\k\w\quad\text{s.t.}\quad \w^H\a^\k=1.
\end{equation}
${\bf C}_{\bf y}^\k={\rm E}[{\bf y}^\k({\bf y}^\k)^H]$ and ${\bf C}_{\bf x}^\k={\rm E}[{\bf x}^\k({\bf x}^\k)^H]$ denotes the covariance matrix of ${\bf y}^\k$ and ${\bf x}^\k$, respectively; ${\rm E}[\cdot]$ is the expectation operator. Hence, $\w^H{\bf C}_{\bf y}^\k\w$ and  $\w^H{\bf C}_{\bf x}^\k\w$ is the power of the residual noise and of the overall output in ${\bf w}^H\x^\k$, respectively. The distortionless constraint $\w^H\a^\k=1$ ensures that the SOI remains untouched in the beamformer's output. For nonsingular covariance matrices, it holds that
\begin{align}
    \w_{\rm MVDR}^\k&=\frac{({\bf C}_{\bf y}^\k)^{-1}\a^\k}{(\a^\k)^H({\bf C}_{\bf y}^\k)^{-1}\a^\k}\label{MVDR},\\
    \w_{\rm MPDR}^\k&=\frac{({\bf C}_{\bf x}^\k)^{-1}\a^\k}{(\a^\k)^H({\bf C}_{\bf x}^\k)^{-1}\a^\k}\label{MPDR}.
\end{align}

Practical deployments require knowledge of $\a^\k$ and of the covariance matrices. MPDR is known to be sensitive to errors in the estimates of ${\bf C}_{\bf x}^\k$ and $\a^\k$ \cite{vantrees2002}. MVDR is less sensitive but it requires knowledge of ${\bf C}_{\bf y}^\k$ or its estimate. 

\subsection{Independent Vector Extraction}
IVE solves the extraction (and localization) problem blindly by jointly estimating $\a^\k$ and $\w^\k$ using only the observed signals. 
The mixture \eqref{BSEmodel} is assumed to be determined\footnote{The determined case is convenient in terms of mathematical feasibility and is commonly used in practice where this condition is not satisfied.}, which means that ${\bf y}^\k$ are assumed to belong to a $(d-1)$-dimensional subspace of so-called background sources represented by the $(d-1)\times 1$ vector ${\bf z}^\k$. The {\em main} assumption is that $s^\k$ is independent of ${\bf y}^\k$ resp. ${\bf z}^\k$, following the ICA model. The dependencies between the elements of the vector component $\s=[s^{[1]},\dots,s^{[K]}]^T$ are taken into account, following the IVA model \cite{kim2006}.

The higher-order statistics-based source model assumes that the samples of signals are i.i.d. (independently and identically distributed); namely, $\s$ is distributed according to a multivariate non-Gaussian density, involving the dependencies among its elements. The pdf of ${\bf z}^\k$ can be assumed circular Gaussian and uncorrelated across mixtures. This source model often neglects useful features of signals, nevertheless, it is simple and guarantees the identifiability of $\{\w^\k,\a^\k\}_{k=1}^K$ \cite{anderson2014,kautsky2019icassp}. 

Having $N$ samples of observations for each $k$, the (quasi) maximum likelihood estimates $\{\hat\w^\k,\hat\a^\k\}_{k=1}^K$ can be obtained by finding the corresponding local maximum of 
\begin{multline}\label{contrast}
    \mathcal{C}\left(\{\hat\w^\k,\hat\a^\k\}_{k=1}^K\right) =\hat{\rm E}\left[\log f\left(\left\{\frac{\hat{s}^\k}{\hat\sigma_k}\right\}_{k=1}^K\right)\right]  \\ -\sum_{k=1}^{K}\log \hat\sigma_k^2 -\sum_{k=1}^{K} \hat{\rm E}\left[(\hat\z^\k)^H({\bf C}_{\bf z}^\k)^{-1}\hat{\bf z}^\k\right]  \\ 
    + (d-2)\sum_{k=1}^{K}\log|\gamma^\k|^2 + \text{const.},
\end{multline}
see, e.g., \cite{koldovsky2021fastdiva}. Here, $f(\cdot)$ is the model pdf of the SOI (a surrogate for the unknown true pdf) corresponding to a normalized random variable; $\hat s^\k=(\hat\w^\k)^H\x^\k$ is the extracted SOI given the current estimate of $\w^\k$; $\hat\sigma_k^2=(\hat\w^\k)^H\widehat{\bf C}_\x^\k\hat\w^\k$ is the sample-based variance of $\hat s^\k$; $\hat{\bf z}^\k=\hat{\bf B}^\k\x^\k$ are background signals estimates through the blocking matrix $\hat{\bf B}^\k=[\hat\g^\k\quad -\hat\gamma^\k{\bf I}_{d-1}]$ where $\hat\a^\k=\left[\begin{smallmatrix}\hat\gamma^\k\\\hat\g^\k\end{smallmatrix}\right]$; $\hat{\rm E}[\cdot]$ denotes the sample-average operator; ${\bf C}_{\bf z}^\k$ is the covariance matrix of ${\bf z}^\k$, which is not known and must be replaced, e.g., by $\widehat{\bf C}_{\bf z}^\k=\hat{\rm E}[\hat{\bf z}^\k(\hat{\bf z}^\k)^H]$.

IVE involves indeterminacies, which are inherent (not only \cite{wang2018}) to blind methods: The role of the SOI can be played by any independent source in the mixture. It can also happen that a different source component (frequency) is extracted in each mixture, which is referred to as the permutation problem \cite{sawada2004sap}. IVE alleviates this by taking into account the dependencies among the elements of $\s$. However, even if the permutation problem is solved, the global uncertainty of sources (speakers) remains. 


This explains why it is necessary to look for the desired local maximum of \eqref{contrast}. The optimization algorithm must be appropriately initialized and controlled. However, there is no way to ensure this on the basis of current assumptions. Additional information about the SOI is needed, which we assume in the following section. 

\section{Proposed Algorithm}
\label{sec:fastica}

\subsection{Weighted covariance matrix}
Let us assume that information about the SOI is available in the form of scalar signals $r_k(n)$, $k=1,\dots,K$. Typically, these signals can correspond to preliminary SOI estimates, activity indicators, estimated variance profiles and similarly so. The information can be joined for all $k$, in which case we have only one scalar signal $r(n)=r_1(n)=\dots=r_K(n)$.

Then, let weighting functions $\alpha_k(\cdot)$, $k=1,\dots,K$, be given, which should be functions of $r_k(n)$. For simplicity, we will assume that the weighting functions are the same for all $k$, corresponding to the function denoted by $\alpha(\cdot)$. Then, we introduce weighted sample covariance matrices as
\begin{equation}\label{weightedcovariance}
    \widehat{\bf C}_\alpha^\k = \hat{\rm E}\left[\alpha(r_k)\x^\k\bigl(\x^\k\bigr)^H\right], \quad k=1,\dots,K.
\end{equation}

Most typically, $\alpha(\cdot)$ is a non-negative and non-increasing function such as, for example, 
\begin{equation}\label{weightingfunctiontypical}
    \alpha(r_k(n))=\frac{1}{\epsilon+|r_k(n)|^2},
\end{equation}
where $\epsilon>0$ is a small constant that prevents division by zero. If $r_k(n)$ somewhat indicates the activity of the SOI, the function \eqref{weightingfunctiontypical} applied in \eqref{weightedcovariance} is aimed to outweigh samples where the SOI is active (and vice versa) so that $\widehat{\bf C}_\alpha^\k$ is as close as possible to ${\bf C}_{\bf y}^\k$.

\subsection{MVDR constraint}
Couplings between the parametric vectors $\a^\k$ and $\w^\k$ can be applied to speed up or stabilize IVE algorithms. A popular one is the orthogonal constraint (OC), which enforces a zero sample-based correlation between $\hat s_k$ and $\hat\z^\k$. Analytically, the OC is identical to MPDR given by \eqref{MPDR}, see \cite{koldovsky2019TSP}. So if $\hat\w^\k$ is the dependent variable, the OC is given by 
\begin{equation}\label{OC}
  \hat\w_{\rm OC}^\k=\frac{(\widehat{\bf C}_{\bf x}^\k)^{-1}\hat\a^\k}{(\hat\a^\k)^H(\widehat{\bf C}_{\bf x}^\k)^{-1}\hat\a^\k},
\end{equation}
where $\widehat{\bf C}_{\bf x}^\k=\hat{\rm E}[{\bf x}^\k({\bf x}^\k)^H]$. Therefore, orthogonally constrained IVE algorithms are seeking the optimum of $\mathcal{C}(\{\hat\w^\k_{\rm OC},\hat\a^\k\}_{k=1}^K)$ with respect to $\{\hat\a^\k\}_{k=1}^K$ and can thus be seen as blind MPDR beamformers \cite{koldovsky2017eusipco}.

Motivated by this interpretation, we propose to replace \eqref{OC} by
\begin{equation}\label{MVDRconstraint}
  \hat\w_{\alpha}^\k=\frac{(\widehat{\bf C}_\alpha^\k)^{-1}\hat\a^\k}{(\hat\a^\k)^H(\widehat{\bf C}_\alpha^\k)^{-1}\hat\a^\k}.
\end{equation}
Provided that $\widehat{\bf C}_\alpha^\k$ is an estimate of ${\bf C}_{\bf y}^\k$, then \eqref{MVDRconstraint} could be seen as an estimate of the MVDR beamformer given by \eqref{MVDR}. 

\subsection{Second-order algorithm}
We now propose an algorithm that 
seeks the optimum point of 
$\mathcal{C}(\{\hat\w^\k_{\alpha},\hat\a^\k\}_{k=1}^K)$ with respect to $\{\hat\a^\k\}_{k=1}^K$, which can be interpreted as a semi-blind MVDR beamformer since $\widehat{\bf C}_\alpha^\k$ provide side information. The optimization approach is heuristically derived based on the approximate Newton-Raphson update \cite{koldovsky2021fastdiva}
\begin{equation}\label{generalupdate}
    \hat\a^\k \leftarrow \hat\a^\k-({\bf H}^{\k*})^{-1}\boldsymbol{\Delta}^\k, \quad k=1,\dots,K,
\end{equation}
where $(\cdot)^{*}$ denotes the conjugate value, and
\begin{align}\label{generalgradient}
    \boldsymbol{\Delta}^\k &= \frac{\partial}{\partial (\a^\k)^*}\mathcal{C}(\{\hat\w^\k_{\alpha},\hat\a^\k\}_{k=1}^K),\\ \label{generalHessian}
    {\bf H}^\k &= \frac{\partial (\boldsymbol{\Delta}^\k)^T}{\partial \a^\k}\Bigg|_{\hat\a^\k=\a^\k,N\rightarrow+\infty}.
\end{align}
After some manipulations, we arrive at the following update rule 
\begin{equation}\label{update}
   \hat\a^\k \leftarrow \hat\a^\k- \frac{\hat\nu_k}{\hat\nu_k-\hat\rho_k}\frac{\hat\sigma^2_{\alpha,k}}{\hat\sigma^2_k}
   \Bigg(
   \hat\a^\k -\hat\nu_k^{-1}\hat{\rm E}\Bigg[\phi_k\left(\overline{\bf s}\right)\frac{\x^\k}{\hat\sigma_k}\Bigg]
   \Bigg),
\end{equation}
where $\phi_k=-\frac{\partial}{\partial s_k}\log f$, $\hat\nu_k =\hat{\rm E}\left[\frac{\hat s_k}{\hat\sigma_k} \phi_k\left(\overline{\bf s}\right)\right]$, $\hat\rho_k =\hat{\rm E}\left[\frac{\partial \phi_k}{\partial s_k^*} \left(\overline{\bf s}\right)\right]$, $\hat\sigma^2_{\alpha,k}=(\hat\w_{\alpha}^\k)^H\widehat{\bf C}_\alpha^\k\hat\w_{\alpha}^\k$, and $\overline{\bf s}=[\frac{\hat s_1}{\hat\sigma_1},\dots,\frac{\hat s_K}{\hat\sigma_K}]^T$ is the normalized SOI vector component.
Detailed derivations are not given in this paper due to limited space; see the Appendix.


\subsection{Relation to the FastICA algorithm}
Let us consider the case $K=1$, in which case the index $k$ can be omitted from the notation. We can easily represent a situation where $r(n)$ does not bring any useful information, e.g., by putting it equal to a constant such that $\alpha=1$. Then, $\widehat{\bf C}_\alpha=\widehat{\bf C}_{\bf x}$ and $\hat\sigma^2_{\alpha}=\hat\sigma^2$. We can assume that the output signal is normalized so that $\hat\sigma^2=1$. Then, \eqref{update} simplifies to 
\begin{equation}\label{updatefastica}
    \hat\a \leftarrow \hat\a- \frac{\hat\nu}{\hat\nu-\hat\rho}
   \left(\hat\a -\hat\nu^{-1}\hat{\rm E}[\phi(\hat s)\x]\right).
\end{equation}
According to the constraint \eqref{MVDRconstraint}, which now coincides with \eqref{OC}, we can apply substitution $\hat\a=\widehat{\bf C}_{\bf x}\hat\w$ and rewrite \eqref{updatefastica} to the form
\begin{equation}\label{updatefastica2}
    \hat\w \leftarrow \hat\w - \frac{\hat\nu}{\hat\nu-\hat\rho}
    \left(\hat\w -\hat\nu^{-1}\widehat{\bf C}_{\bf x}^{-1}\hat{\rm E}[\phi(\hat s)\x]\right).
\end{equation}
Since the scale of the output signal (thus also of the parameter vector $\w$) can be arbitrary, the right-hand side of \eqref{updatefastica2} can be multiplied by the scalar $\hat\rho-\hat\nu$, which results in the well known FastICA update rule \cite{hyvarinen1999}
\begin{equation}
   \hat\w \leftarrow \hat\rho\hat\w - \widehat{\bf C}_{\bf x}^{-1}\hat{\rm E}[\phi(\hat s)\x].
\end{equation}
The update \eqref{update} thus can be seen as an informed extension of the FastICA algorithm.

\section{Experimental validation}
\label{sec:experiments}
\subsection{Simulations}
In simulations, we generate artificial mixtures and extract the SOI to verify the performance of the proposed algorithm and compare it with the original one-unit FastICA\footnote{We use the implementation from \cite{koldovsky2022double} with $K=T=L=1$.}. The algorithms are compared in two regimes: (1) mixtures are processed separately as if $K=1$ (the original and proposed method are denoted by ``FastICA'' and ``iFastICA", respectively) and jointly as if $K=6$ (the methods are denoted ``FastIVA'' and ``iFastIVA", respectively). We primarily focus on their global convergence, i.e., whether the algorithms extracted the desired source. Each extracted signal is evaluated by means of signal-to-interference ratio (SIR); the success rate of an algorithm is defined as the percentage of trials where it achieves SIR higher than $3$dB. The SIR is then averaged over these ``successful'' trials, which reflects the accuracy of the algorithm.

In a trial of the simulation, a complex-valued mixture of $d=5$ super-Gaussian signals of length $N$ is generated. The first column of the random mixing matrix is the mixing vector of the SOI, which is the first generated signal. The initial SIR (SIR$_{\rm ini}$), defined as the ratio of its scale to the mean scale of the other signals, is set to the chosen value through a corresponding re-scaling of the SOI. The side information in iFastICA/iFastIVA is provided through $r_k(n)=\sqrt{1-\epsilon^2}s^\k(n)+\epsilon v^\k(n)$, where $v^\k(n)$ is standard Gaussian, and $\epsilon^2=0.5$; the weighting function is given by \eqref{weightingfunctiontypical}. The algorithms are initialized randomly in a random vicinity of the SOI.

\begin{figure}[ht]
    \centering
    \includegraphics[width=0.9\linewidth]{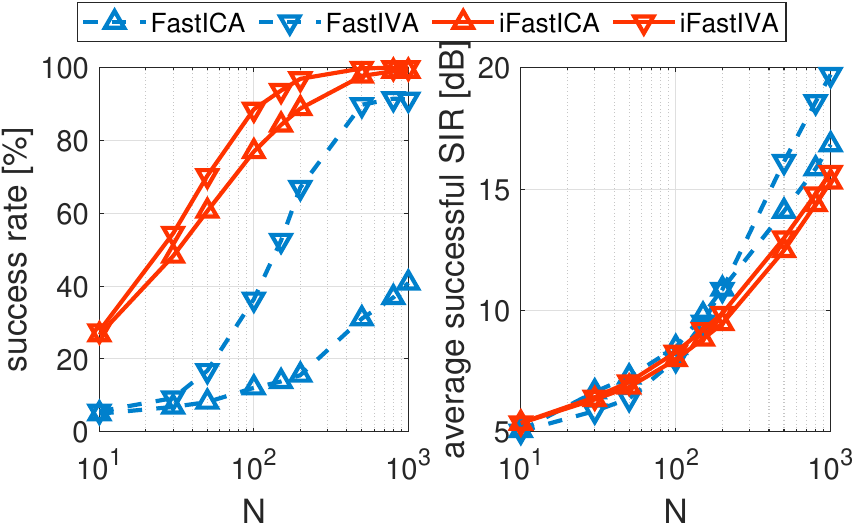}\vspace{\floatsep}\\
    \includegraphics[width=0.9\linewidth]{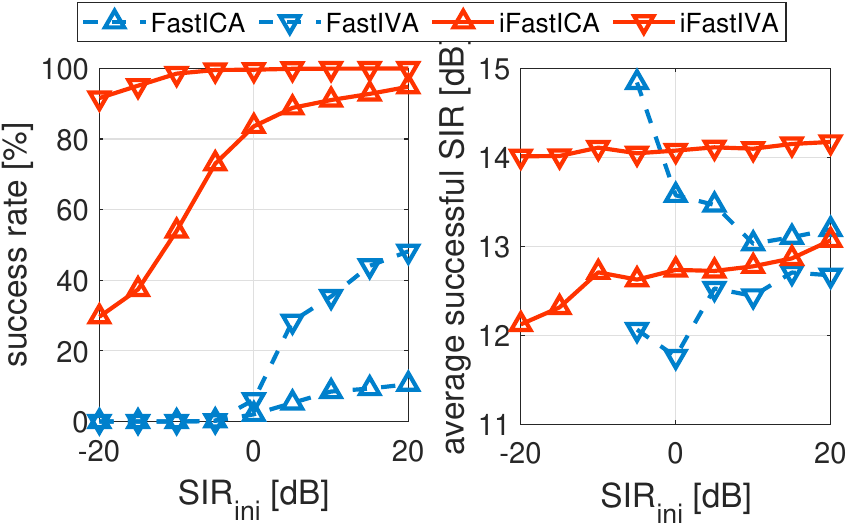}
    \caption{Success rate and SIR averaged over successful trials of the compared algorithms as functions of $N$ (when SIR$_{\rm ini}=0$~dB) and SIR$_{\rm ini}$ (when $N=200$); each setting was repeated in 1000 trials.}
    \label{fig:simulations}
\end{figure}

Fig.~\ref{fig:simulations} shows the success rate and average SIR as functions of $N$ (when SIR$_{\rm ini}=0$~dB) and SIR$_{\rm ini}$ (when $N=200$). The success rate shows that the proposed semi-blind methods significantly profit from the side information compared to the blind algorithms. The accuracy of the semi-blind methods is, on the other hand, similar to that of the blind ones or even slightly worse. The semi-blind methods thus bring improvements in terms of global convergence. 

Of particular interest are the challenging values of $N=10,\dots,100$ and SIR$_{\rm ini}=-20,\dots,-5$~dB. The success rate of blind algorithms here is often less than 10\%, pointing to the problem of blind extraction if the data is too short or if the SOI is too weak. The semi-blind methods offer a solution in this regard. Finally, the performance of iFastIVA/FastIVA is higher than that of iFastICA/FastICA, which clearly demonstrates the benefit of joint extraction, also in the case of the semi-blind methods. 

Simulations can be used to test the algorithms depending on other key parameters such as $\epsilon^2$. The code is publicly available\footnote{\sf https://github.com/koldovsk/IWAENC-2024-Informed-FastICA}.

\subsection{Target speaker extraction}

FastIVA and iFastIVA are compared to the informed auxiliary function-based iAuxIVE (alternative naming for the piloted AuxIVE from~\cite{jansky2022}) in the frequency-domain speaker extraction problem. Reverberated mixtures of two speakers contained in the multi-channel Wall Street Journal dataset (MC-WSJ0-2mix, \cite{mc-wsj0-2mix}) are considered. The dataset contains $3,000$ simulated mixtures; we use the first four microphones. Each mixture contains two active speakers; thus, there are $6,000$ extraction experiments in total. The sources are mixed at SIR between $\langle-5,+5\rangle$~dB. The recordings are highly reverberant ($T_{60} \in \langle200,600\rangle$~ms).
The sampling frequency is $16$~kHz; the STFT window length is 1000 samples; the window shift is 200. 

The prior information is obtained using speaker identification via embeddings. The embeddings are computed in the same manner as described for the MC-WSJ0-2mix in~\cite{loc:bse:defl}. The signal $r(n)$ (independent of $k$) and the pilot signal for the iAuxIVE are equal to the energy of the mixture when the target speaker appears to be dominant and zero otherwise. The target is assumed dominant within a closed set of speakers when its reference embedding is the most similar to the embedding computed using the $n$th frame of the mixture.
%

\begin{figure}
     \centering
     \includegraphics[width=0.42\linewidth]{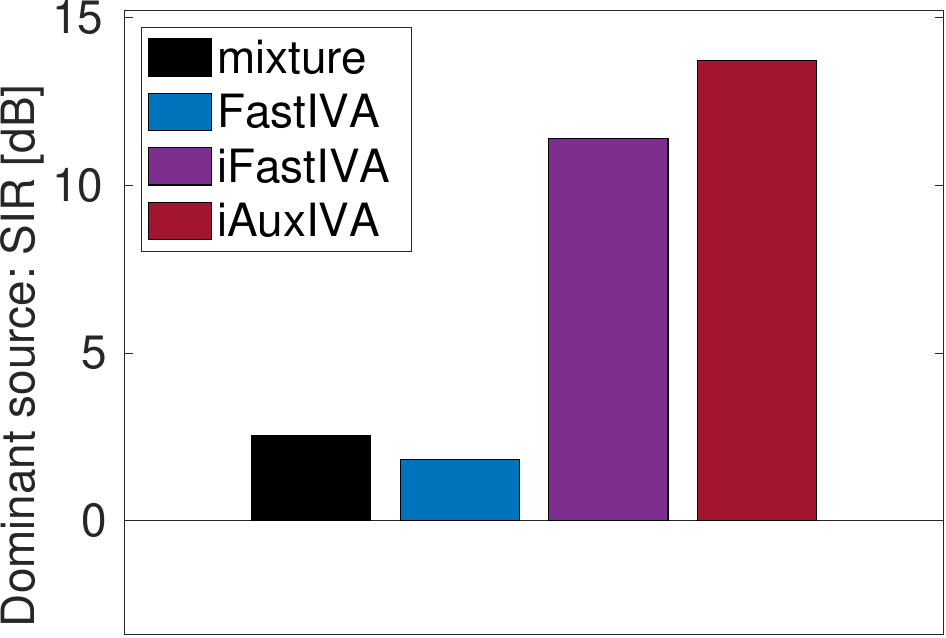}\hfill
     \includegraphics[width=0.42\linewidth]{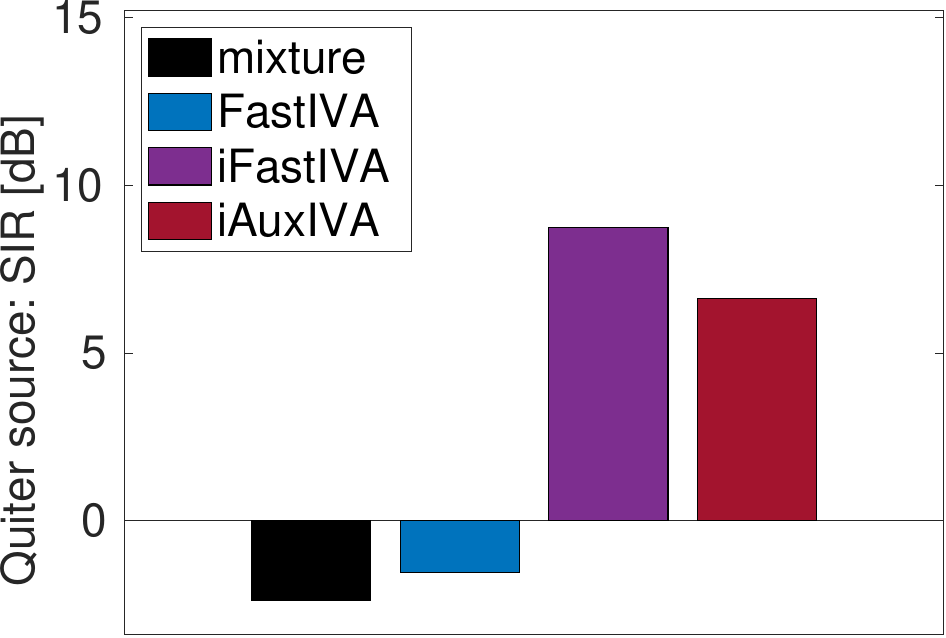}\\     
     \includegraphics[width=0.42\linewidth]{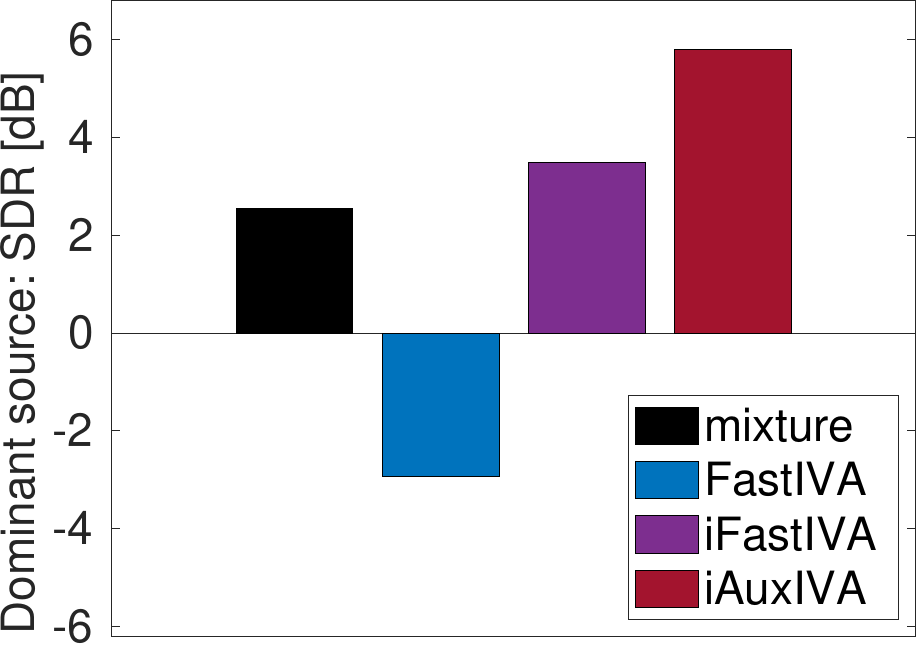}\hfill
     \includegraphics[width=0.42\linewidth]{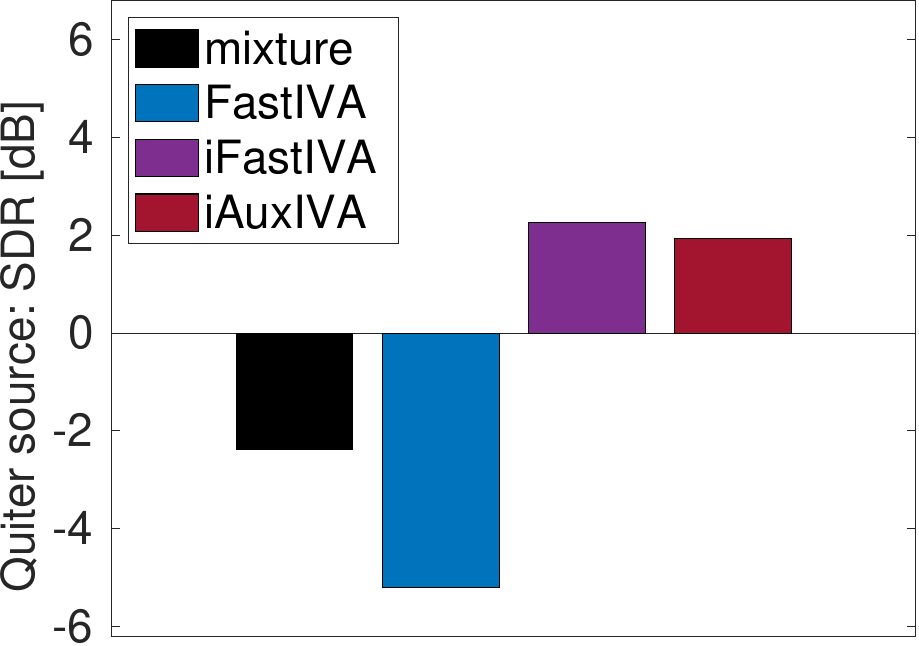}     
     \caption{\label{fig:extr} MC-WSJ0-2mix: Speaker extraction metrics achieved by the compared algorithms.}
     \label{fig:epsilon}
 \end{figure}

Fig.~\ref{fig:extr} presents the results evaluated using Signal-To-Interference Ratio (SIR) and Signal-To-Distortion Ratio (SDR) defined by the BSS\_EVAL~\cite{bsseval} toolbox. The experiment is evaluated separately for the dominant and the quieter speaker. FastIVA fails here as it cannot distinguish the target, whereas both informed algorithms are able to extract it. The iAuxIVE algorithm is more successful when extracting the dominant source, whereas iFastIVA excels at extracting the quieter speaker.

\section{Conclusions}
\label{sec:conclusions}
We have proposed semi-blind algorithms for source extraction that can exploit side information to reach the desired source. The algorithms combine IVE and MVDR intuitively. Future work should focus on deeper theoretical analysis to provide insight into which type of side information can be used most effectively.

\section{Appendix}
The computation of \eqref{generalgradient}, after $\phi_k$ is divided by $\hat\nu_k$ (see \cite{koldovsky2021fastdiva} for justification of this step), gives
\begin{multline}\label{gradient}
   \boldsymbol{\Delta}^\k = \sigma^2_{\alpha,k}(\widehat{\bf C}_\alpha^\k)^{-1}\Bigg(
   \hat\a^\k -\hat\nu_k^{-1}\hat{\rm E}\Bigg[\phi_k(\overline{\bf s})\cdot\frac{\x^\k}{\sigma_k}\Bigg]
   \Bigg)+\\
   \begin{pmatrix}
   (\hat{\bf g}^\k)^H(\widehat{\bf C}_{\bf z}^\k)^{-1}{\bf q}^\k\\
   -(\hat\gamma^\k)^*(\widehat{\bf C}_{\bf z}^\k)^{-1}{\bf q}^\k
   \end{pmatrix},
\end{multline}
where ${\bf q}^\k=\hat{\rm E}[\hat\z^\k\hat s^\k]$, which is the sample covariance of the current SOI estimate and the background signals. For the orthogonal constraint \eqref{OC}, ${\bf q}^\k$ equals zero. To ensure this, $\hat\a^\k$ should be recomputed after \eqref{MVDRconstraint} as $\hat\a^\k\leftarrow \hat\sigma_k^{-2}\widehat{\bf C}_{\bf x}^\k\hat\w^\k$, which is equivalent to \eqref{MPDR}. 
Next, in the computation of \eqref{generalHessian}, we neglect all rank-one terms, which results in
\begin{equation}\label{hessian}
    {\bf H}^\k=\hat\sigma_{\alpha,k}^2\left({\bf I}_d-\frac{\hat\rho_k}{\hat\nu_k}\frac{\hat\sigma^2_{\alpha,k}}{\hat\sigma^2_k}(\widehat{\bf C}_\alpha^{\k*})^{-1}
   \widehat{\bf C}_{\bf x}^{\k*} \right)(\widehat{\bf C}_\alpha^{\k*})^{-1}.
\end{equation}
Then, we apply the approximation  $\frac{\hat\sigma^2_{\alpha,k}}{\hat\sigma^2_k}(\widehat{\bf C}_\alpha^{\k})^{-1}
   \widehat{\bf C}_{\bf x}^{\k}\approx {\bf I}_d$ and multiply 
${\bf H}^\k$ by an experimentally verified factor $\frac{\hat\sigma^2_k}{\hat\sigma^2_{\alpha,k}}$. By putting into \eqref{generalupdate}, \eqref{update} follows.\hfill\rule{1.2ex}{1.2ex}

\bibliographystyle{IEEEbib_abb}

\end{document}